\documentclass[10pt]{article} 
\usepackage[preprint]{tmlr}
\usepackage{graphicx} 
\usepackage[hyphens]{url}  
\usepackage{graphicx} 
\usepackage{natbib}  
\usepackage{caption} 
\usepackage{algorithm}
\usepackage{algorithmic}

\usepackage{microtype}      
\usepackage{xcolor}         

\usepackage{hyperref} 
\usepackage{cleveref}
\usepackage{enumitem}

\usepackage{url}            
\usepackage{booktabs}       
\usepackage{amsfonts}       
\usepackage{nicefrac}       

\usepackage{subcaption}
\usepackage[toc,page]{appendix}

\title{Permission Manifests for Web Agents\\[0.8em]{\large Lightweight Agent Standards Working Group (LAS-WG)}\vspace{-1.05em}}

\author{%
  Samuele~Marro$^{1,2}$\thanks{Corresponding author. Email: \href{mailto:samuele.marro@eng.ox.ac.uk}{samuele.marro@eng.ox.ac.uk}},
  Alan Chan$^{3}$,
  Xinxing Ren$^{4}$,
  Lewis Hammond$^{1,5}$,
  Jesse Wright$^{1,2}$,
  Gurjyot Wanga$^{6}$,
  Tiziano Piccardi$^{7}$,
  Nuno Campos$^{8}$,
  Tobin South$^{9,2}$,
  Jialin Yu$^{1,2}$,
  Sunando Sengupta$^{10}$,
  Eric Sommerlade$^{10}$,
  Alex Pentland$^{9}$,
  Philip Torr$^{1,2}$,
  Jiaxin Pei$^{11,9,2}$\\[0.5em]
  {\normalfont
  $^{1}$University of Oxford\quad
  $^{2}$Institute for Decentralized AI\quad
  $^{3}$Centre for the Governance of AI\quad
  $^{4}$Coral Protocol\\
  $^{5}$Cooperative AI Foundation\quad
  $^{6}$Webair\quad
  $^{7}$Johns Hopkins University\quad
  $^{8}$Witan Labs\\
  $^{9}$Stanford University\quad
  $^{10}$Microsoft\quad
  $^{11}$UT Austin
  }
}

\date{August 2025}

\begin{document}

\maketitle


\begin{abstract}
The rise of Large Language Model (LLM)-based web agents represents a significant shift in automated interactions with the web. Unlike traditional crawlers that follow simple conventions, such as \texttt{robots.txt}, modern agents engage with websites in sophisticated ways: navigating complex interfaces, extracting structured information, and completing end-to-end tasks. Existing governance mechanisms were not designed for these capabilities. Without a way to specify what interactions are and are not allowed, website owners increasingly rely on blanket blocking and CAPTCHAs, which undermine beneficial applications such as efficient automation, convenient use of e-commerce services, and accessibility tools.
We introduce \texttt{agent-permissions.json}, a \texttt{robots.txt}-style lightweight manifest where websites specify allowed interactions, complemented by API references where available. This framework provides a low-friction coordination mechanism: website owners only need to write a simple JSON file, while agents can easily parse and automatically implement the manifest's provisions. Website owners can then focus on blocking non-compliant agents, rather than agents as a whole. By extending the spirit of \texttt{robots.txt} to the era of LLM-mediated interaction, and complementing data use initiatives such as AIPref, the manifest establishes a compliance framework that enables beneficial agent interactions while respecting site owners' preferences.
\end{abstract}

\begin{figure}[ht]
  \centering
  \begin{subfigure}{0.45\textwidth}
    \centering
    \includegraphics[width=\linewidth]{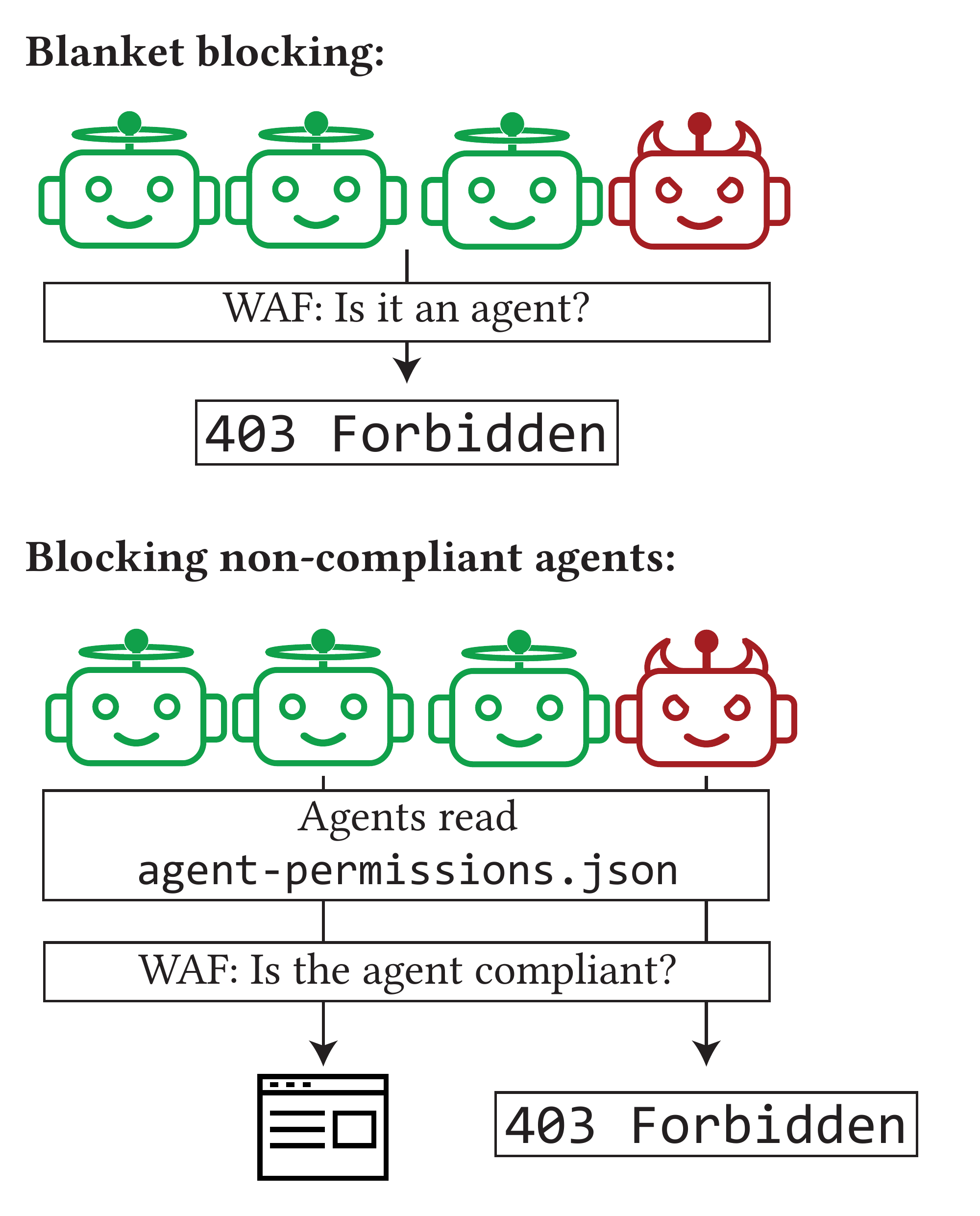}
    \caption{Before and after.}
    \label{fig:abstractA}
  \end{subfigure}
  \begin{subfigure}{0.45\textwidth}
    \centering
    \includegraphics[width=\linewidth]{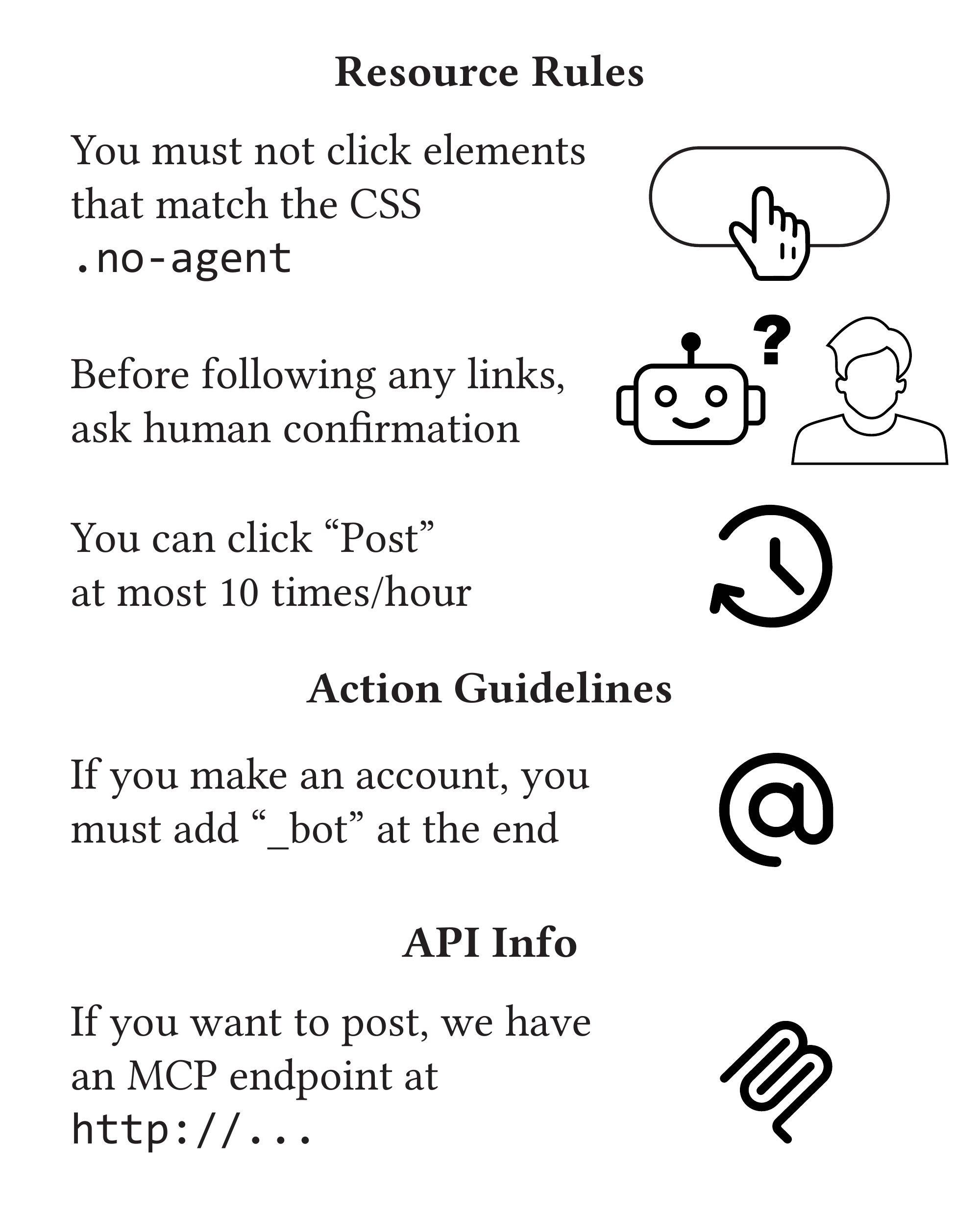}
    \caption{Examples of rules.}
    \label{fig:abstractB}
  \end{subfigure}
  \caption{\texttt{agent-permissions.json} is a minimalist, straightforward permission manifest for agents interacting with web pages.}
  \label{fig:abstract}
\end{figure}

\section{Introduction}
The proliferation of Large Language Model (LLM)-based web agents has fundamentally changed how automated systems interact with web content. Unlike traditional web crawlers that primarily index content, modern LLM-powered agents perform sophisticated interactions: conducting searches to find specific information \citep{nakano2021webgpt}, submitting forms as part of automated workflows \citep{drouin2024workarena,boisvert2024workarenapp}, and executing end-to-end tasks \citep{wang2024oscar,erdogan2025plan}. These capabilities enable powerful applications, such as intelligent research assistants \citep{baek2024researchagent} and automated accessibility tools \citep{kodandaram2024enabling}, as well as new opportunities for e-commerce \citep{zhang2024llasa} and recommendation systems \citep{lazar2025moral}.

However, even beneficial tools can overwhelm websites or violate usage policies when deployed at scale \citep[see, e.g.,][]{bort2025how}. 
Unable to distinguish between beneficial AI agents and potentially harmful automated traffic \citep{corral2025perplexity,reuters2025reddit}, many site operators have adopted a defensive approach, implementing blanket 403 (``Forbidden'') responses, CAPTCHA challenges, and aggressive rate limiting against any client exhibiting suspected automated behavior \citep{fletcher2024many,llamas2025balancing,iliou2021detection,bocharov2024declare}. 
While understandable from a security perspective, this approach creates significant barriers to legitimate AI-driven services \citep{w3c2021inaccessibility}, depriving both users and websites of the value of service automation.  

A key bottleneck is the lack of a standardized and lightweight mechanism for website owners to express nuanced policies: which automated interactions are welcome, which are prohibited, and under what conditions different behavior are allowed. For example, a travel-booking assistant might be allowed to check seat availability or flight status, but should not confirm or pay for bookings without human approval. Similarly, an accessibility auditor might be allowed to activate dropdown menus and form fields to verify usability, but might be forbidden from submitting or modifying data.
Such policies cannot currently be expressed in a standardized fashion, which means that website owners face a binary choice: either block all automated access, or implement complex, burdensome authentication systems requiring API keys, OAuth flows, and extensive integration efforts. For the majority of the web, where ambiguity (rather than malice) is a major problem, a simple coordination mechanism could help substantially.

To address this problem, this paper proposes \texttt{agent-permissions.json}, a \texttt{robots.txt}-style permission manifest\footnote{\url{https://github.com/las-wg/agent-permissions.json}}. 
Drawing inspiration from the simplicity and widespread adoption of the \texttt{robots.txt} standard, we present a lightweight framework that allows website owners to declaratively specify interaction policies for AI agents (\Cref{fig:abstractB}). This approach requires no credential provisioning, complex authentication infrastructure, or ongoing maintenance overhead: it is merely a simple manifest file that agents can easily discover and follow.

Our thesis is that a permission manifest provides a familiar, low-friction entry point for signaling permissible interactions with a website. Rather than treating all agents as a threat, we argue that the community should make a distinction: some agents are \emph{compliant}, i.e., open to following website policies when performing actions on behalf of the user, while others are \emph{rule-breaking}. We suspect that many agents would be compliant if given the opportunity, just as many major search crawlers comply with \texttt{robots.txt}. \texttt{agent-permissions.json} is meant to coordinate compliant agents so that they can both respect website policies and provide useful services for the user. Rule-breaking agents will remain a problem, but websites can focus their detection and blocking efforts on them while benefiting from the traffic of compliant agents. Not all websites will want to make this trade-off, especially those whose funding depends upon human attention (e.g., from ads). Our primary target is sites whose value comes from delivering information and services, such as e-commerce platforms, documentation and research repositories, and scheduling and booking interfaces. Such websites will likely find \texttt{agent-permissions.json} attractive.  


Our proposed manifest, coupled with data use standards (e.g., AIPref\footnote{\url{https://datatracker.ietf.org/wg/aipref/about/}}), aims to preserve the autonomy of website owners while creating predictable interaction patterns that benefit developers, end users, and the broader ecosystem of AI-powered web services.


\section{Background and Positioning}
\label{sec:background}

To understand the need for a permission manifest system for AI agents, we must first examine the historical context of web automation governance and the limitations of current approaches. 
We thus provide a quick summary of existing standards and how they fit different use cases in \Cref{tab:useCases}.

\begin{table}[h]
  \centering
  \begin{tabular}{@{}ll@{}}
    \toprule
    \textbf{Use Case} & \textbf{Solution} \\
    \midrule
    Permissions for crawling & \texttt{robots.txt} \\
    Guiding agentic crawlers & \texttt{llms.txt} \\
    Permissions for data use & AIPref, \texttt{ai.txt} \\
    Offering structured access & REST, MCP, A2A \\
    Blocking malicious agents & WAF, CAPTCHA \\
    Monetizing agentic interactions & Pay per Crawl, x402, AP2 \\
    Permissions for UI interactions & \texttt{agent-permissions.json} (\textbf{Ours}) \\
    \bottomrule
  \end{tabular}
  \caption{Existing and proposed web governance mechanisms for agents, and where \texttt{agent-permissions.json} fits.}
  \label{tab:useCases}
\end{table}

\paragraph{robots.txt.}
The \texttt{robots.txt} standard \citep{rfc9309}, introduced in 1994, represents one of the web's most successful examples of lightweight, declarative access control. Site owners communicate crawling preferences to automated agents through a standardized text file placed at the root of their website.

The protocol's widespread adoption stems from two key characteristics that make it an ideal model for modern web automation governance.
First, \texttt{robots.txt} requires minimal technical infrastructure. Website operators only need to create a plain text file with straightforward syntax: no database configurations, authentication servers, or complex middleware deployments. This simplicity has made \texttt{robots.txt} the de-facto standard for crawl management on the Web \citep{robots2025}. Second, the standard places minimal burden on agents. Web crawlers have universally implemented \texttt{robots.txt} parsing, making compliance a solved problem rather than a recurring engineering challenge.

The \texttt{robots.txt} approach also demonstrates the effectiveness (and limitations) of voluntary compliance in web governance. Although nothing technically prevents a crawler from ignoring \texttt{robots.txt} directives \citep{kim2025scrapers,paul2024multiple}, the major crawling organizations (particularly search engines) have consistently followed these rules \citep{cloudflareWebCrawler}, making the negative impact of rule-breakers tolerable. This success suggests that, given sufficient adoption by major players, declarative permission systems can work when the incentives of both publishers and consumers of web content are sufficiently aligned. For a counter-example of this phenomenon, consider instead web scrapers: since their objective is to gather as much data as possible without regards for the website's intentions, \texttt{robots.txt} is often ignored \citep{cloudflareWebCrawler}. The key takeaway is that, while permission manifests cannot align incentives on their own, they provide a framework through which aligned entities can coordinate.


\paragraph{The Emerging Gap.}
Despite the benefits of \texttt{robots.txt}, a significant gap exists between the simple crawling paradigm that \texttt{robots.txt} was designed to address and the rich interaction patterns that modern AI agents require. Traditional web crawling involves relatively straightforward operations: fetching HTML documents, following links, and building indexes. The \texttt{robots.txt} standard adequately covers these use cases through its allow/disallow directives and crawl-delay parameters (though the latter are not always supported\footnote{\url{https://developers.google.com/search/docs/crawling-indexing/robots/robots_txt}}).

Contemporary LLM-driven agents, however, engage in fundamentally different interaction patterns. They may perform semantic queries that extract specific information from dynamic content, submit forms to test workflows or gather data, interact with JavaScript-rendered elements, or transform content in real-time for summarization or translation \citep{zhou2023webarena}. \texttt{robots.txt} lacks the expressivity to encode nuanced policies about how and when AI interactions are permitted \citep{reitinger2025measured,li2025ai}.\footnote{An attempt has been made to use \texttt{robots.txt} as an informal signaling mechanism for opting out of AI training and AI-mediated interactions \citep{pierce2024text}, though this approach does not allow any nuance in data use or allowed interactions.}

This gap has created a situation where valuable AI services are increasingly blocked by defensive measures designed to combat indiscriminate scraping. The result is a degraded ecosystem where beneficial automation is often indistinguishable from potentially harmful activity, leading to broad restrictions that limit beneficial uses of AI while failing to effectively address security concerns. The rest of this section discusses modern approaches to automated interaction management.

\paragraph{Heavyweight Alternatives and Their Limitations.}
Contemporary approaches to web automation control typically involve significantly more complex infrastructure. OAuth 2.0 flows, while robust and secure, require website operators to implement authorization servers, manage client credentials, and handle token lifecycle management \citep{rfc6749}.
API key systems demand similar infrastructure plus ongoing operational workflows: provisioning keys for each app/user, scoping and restricting them (by referrer, IP/app, and API), storing and distributing keys securely, monitoring usage and quotas, and regularly rotating or revoking keys on leakage or churn \citep{nistsp80057,googlecloudApi,owaspfoundation2025key}.

Web Application Firewalls (WAFs) and bot detectors typically operate through pattern matching and behavioral analysis rather than explicit permission declaration \citep{prandl2015study}. While effective at blocking malicious traffic, WAFs provide no standard mechanism for site owners to communicate allowed interactions, forcing agents to guess or rely on circumstantial information. In other words, they can only be used as enforcement systems, not as coordination mechanisms. 

These heavyweight solutions share common drawbacks that limit their applicability to the broader web ecosystem. They require significant upfront engineering investment, ongoing operational overhead, and specialized security expertise \citep{sun2012devil,yang2016model}. For the vast majority of websites, particularly smaller sites, personal blogs, and content-focused platforms, such systems likely require prohibitive complexity relative to their benefits.

Moreover, these approaches create barriers to beneficial automation. For example, a researcher building an agent-based accessibility tool or a developer creating an e-commerce assistant agent will face complex integration requirements. This high barrier to entry often forces legitimate use cases to employ workarounds, such as screen scraping or reverse engineering \citep{olston2010web,trezza2023scrape,brown2024web}, ultimately degrading the quality and reliability of beneficial automated services.

\paragraph{llms.txt, AIPref, TDMRep, and ai.txt.}
Several recent efforts have sought to define machine-readable standards through which websites can communicate preferences or guidance to AI systems. While these efforts represent important progress, they target distinct layers of the broader coordination problem and leave critical gaps in the governance of automated \emph{interaction}.

The \texttt{llms.txt} proposal \citep{howard2024llmstxt} represents one of the most successful standards for modern LLM signaling. It provides a Markdown-based convention for exposing LLM-friendly content summaries and links, which effectively represent a curated table of contents for language models. An \texttt{llms.txt} file contains brief project descriptions and pointers to Markdown versions of key pages (e.g., \texttt{.html.md}) to simplify retrieval and context construction. This makes \texttt{llms.txt} valuable as an accessibility and comprehension aid, but it lacks any mechanism for expressing or enforcing behavioral constraints. In other words, \texttt{llms.txt} helps agents \emph{read} a site efficiently, but not \emph{interact} with it responsibly.

By contrast, AIPref and TDMRep\footnote{\url{https://www.w3.org/groups/cg/tdmrep/}} (two emerging data-use preference signals) focus explicitly on \emph{content policy}: whether and how publishers permit model training, storage, derivative works, or summarization. This is complementary to our goal. While AIPref and TDMRep aspire to clarify licensing-like semantics for \emph{data} collected by scraping the website, \texttt{agent-permissions.json} specifies permissions for \emph{interactions} with web UIs: what may be clicked or submitted, at what rates, with which safeguards, and when an API should be preferred. We therefore treat AIPref and TDMRep as an adjacent layer in a broader stack: content-use signals govern what models may do with data, while our permission manifest governs how agents may behave on pages.

Finally, \texttt{ai.txt} \citep{li2025ai} introduces a \texttt{robots.txt}-inspired, line-based DSL that lets sites attach per-path and per-element rules to high-level ``AI actions'' (e.g., \texttt{Train}, \texttt{Index}, \texttt{Summarize}). This represents a useful step toward harmonizing how AI systems interpret site intent, but it primarly governs how data from UI elements is processed, rather than how agents interact with the UI itself. For example, \texttt{ai.txt} cannot restrict the clicking of specific HTML elements, impose rate limits, or require human approval before executing actions. As a consequence, it aligns more closely with AIPref and \texttt{robots.txt}  in addressing content-use policies, rather than the UI interaction governance that \texttt{agent-permissions.json} provides.


\paragraph{Pay per Crawl.}
Another emerging approach treats web access as an economic transaction. The concept of micro-payments for content dates to the early web: HTTP status code 402 (``Payment Required'') was reserved in 1992 for this purpose but never gained traction as advertising and subscriptions became the dominant revenue models. Modern AI scrapers, however, threaten these models by consuming content without viewing ads or generating subscriptions, prompting renewed interest in pay-per-access schemes. Cloudflare's Pay per Crawl \citep{cloudflare2025paypercrawl} is an example of this approach: when a recognized crawler requests a page, the site responds with ``payment required'' metadata unless the crawler authenticates and agrees to pay a toll, after which a broker mediates identity, pricing, and settlement.


Compared to Pay per Crawl, \texttt{agent-permissions.json} addresses a different problem: specifying permissible interactions rather than monetizing access. Pay per Crawl systems address only the question of whether a crawler may access a resource contingent on payment. They do not specify the interaction semantics, such as what UI interactions are allowed, or whether they require human approval. The two approaches are compatible, since a website could enforce payment for bulk content access while using a manifest to declare which interactive behaviors are allowed. Critically, \texttt{agent-permissions.json} requires no authorization infrastructure or dependence on a central platform, making it accessible to the long tail of websites that lack the resources or business model for pay-per-access systems.

\paragraph{MCP and A2A.} The Model Context Protocol \citep[MCP;][]{mcp} and Agent-to-Agent (A2A) protocol \citep{a2a} represent complementary approaches to standardizing agent interoperability. MCP provides a framework for agents to discover and invoke external tools through authenticated, schema-based interfaces, while A2A enables direct agent-to-agent communication and capability delegation. Both protocols operate through structured APIs (with added natural language capabilities) rather than unstructured web interfaces, making them closer to REST endpoints than traditional DOM-based interaction. They can be coupled with x402 \citep{x402} and AP2 \citep{ap2}, respectively, to handle payments.
Despite their API-centric design, these protocols remain important components of the web interoperability layer. MCP, in particular, has seen growing integration with web-based systems. For example MCP-B\footnote{\url{https://mcp-b.ai}} enables MCP servers to be embedded directly into web pages and Claude for Chrome interacts with the browser through MCP. This trend suggests that the boundary between API-based and web-based interaction is increasingly fluid, with agents needing to navigate both paradigms.

\texttt{agent-permissions.json} provides native support for advertising MCP and A2A endpoints alongside OpenAPI specifications (see \Cref{sec:api}). This unified discovery mechanism allows website operators to signal their preferred interaction mode (whether it is traditional APIs, MCP tools, A2A agents, or DOM-level automation).
By treating these protocols as first-class alternatives to web scraping, the manifest encourages agents to use structured, stable interfaces when available, while preserving fallback paths for sites that lack such infrastructure.



\section{The Manifest}

Building on the preceding review of crawling-era standards (e.g., \texttt{robots.txt}), heavyweight controls (OAuth, API keys, WAFs), and recent signaling initiatives (\texttt{llms.txt}, AIPref, TDMRep, \texttt{ai.txt}), we argue that modern web agents require a different primitive. The relevant question is not whether a client is automated, but whether its \emph{interactions} conform to site-specific constraints.

We therefore propose \texttt{agent-permissions.json}, a manifest file that governs how agents may interact with web pages, what may be clicked or submitted, at what cadence and concurrency, when human-in-the-loop confirmation is required, and where APIs should be preferred. This section outlines the manifest's key features and their rationale: the distinction between resource-level and action-level permissions, the representation of resource rules and action guidelines, and integration with API references. See \Cref{fig:apf} for examples of rules.

\begin{figure}[t]
  \centering

  \begin{subfigure}[t]{\textwidth}
    \centering
    \fontsize{8}{10}\selectfont\ttfamily

    \begin{minipage}[t]{0.32\linewidth}
\begin{verbatim}
{
  "verb": "click_element",
  "selector": ".no-agent",
  "allowed": false
},
\end{verbatim}
    \end{minipage}\hfill
    \begin{minipage}[t]{0.32\linewidth}
\begin{verbatim}
{
  "verb": "follow_link",
  "selector": "*",
  "allowed": true,
  "modifiers": {
    "human_in_the_loop": true
  }
},
\end{verbatim}
    \end{minipage}\hfill
    \begin{minipage}[t]{0.32\linewidth}
\begin{verbatim}
{
  "verb": "click_element",
  "selector": "#post",
  "allowed": true,
  "modifiers": {
    "rate_limit": {
      "max_requests": 10,
      "window_seconds": 3600
    }
  }
}
\end{verbatim}
    \end{minipage}

    \caption{Resource rules.}
    \label{fig:apf-resource-cols}
  \end{subfigure}

  \vspace{0.8em}

  \begin{subfigure}[t]{0.48\textwidth}
    \centering
    \fontsize{8}{10}\selectfont\ttfamily
\begin{verbatim}
{
  "directive": "MUST",
  "description":
    "Append \"_bot\" to the end of the
     username when creating an account."
},
{
  "directive": "MUST NOT",
  "description":
    "Send direct messages to users.",
  "exceptions":
    "MAY message site administrators."
}



\end{verbatim}
    \caption{Action guidelines.}
    \label{fig:apf-actions-half}
  \end{subfigure}
  \hfill
  \begin{subfigure}[t]{0.48\textwidth}
    \centering
    \fontsize{8}{10}\selectfont\ttfamily
\begin{verbatim}
{
  "type": "openapi",
  "endpoint":
    "https://api.example.com/openapi.yaml",
  "description":
    "Core site API"
},
{
  "type": "mcp",
  "endpoint":
    "mcp://example/agents",
  "docs":
    "https://docs.example.com/mcp",
  "description":
    "Agent task interface"
}
\end{verbatim}
    \caption{API references.}
    \label{fig:apf-api-half}
  \end{subfigure}

  \caption{\texttt{agent-permissions.json} examples: resource rules (top), action guidelines (bottom left), and API references (bottom right).}
  \label{fig:apf}
\end{figure}

\subsection{Structure and Discoverability}

The permission file is a single JSON artifact discoverable in two ways: (i) at a well-known path \texttt{/.well-known/\texttt{agent-permissions.json}}, and (ii) via an HTML link tag \texttt{<link rel="agent-permissions" href="…">} that overrides the root permission file. The link method allows webmasters to specify custom rules for a specific page while leaving the root file as the default.
We chose JSON over a domain-specific language file (like the one used in \texttt{robots.txt}) because JSON is universally supported and can therefore be easily parsed by non-agentic, non-custom parsers.

The file contains four top-level fields:
\begin{itemize}
    \item \texttt{metadata} (mandatory): basic information including schema version, last update timestamp, and document author;
    \item \texttt{resource\_rules} (mandatory): rules governing how agents may interact with HTML elements (\Cref{sec:resourceRules}).
    \item \texttt{action\_guidelines}: high-level, natural-language directives for agent behavior when interacting with the page (\Cref{sec:actionGuidelines});
    \item \texttt{api}: links to documentation for API endpoints providing equivalent functionality to the web interface.
\end{itemize}

\subsection{Resource and Action Permissions}

A core design principle of the manifest is that automated interactions operate along two complementary dimensions: interactions with specific \emph{resources} and general \emph{actions} \citep{south2025authenticated}. Resource permissions specify which interface elements (`resources') an agent may interact with, analogous to resource scoping in access-control theory. This type of rules are a staple of traditional permission management rules. The rise of LLM agents, however, has also enabled the creation of action permissions: rules for higher-level behaviors or workflows (`actions'), such as booking flights or sending direct messages. For example, ``you must not click on the `Post' button'' is an example of a resource rule, where the `Post' button is the resource in question, while ``You must not impersonate an admin'' is an example of an action rule, as it refers to a more complex, ambiguous set of behaviors.

Both dimensions are essential. Resource permissions provide machine-enforceable constraints that the agent's interface layer (e.g., a headless browser) can validate deterministically, preventing inadvertent unauthorized interactions such as repeated form submissions or clicks on sensitive elements. Action permissions, by contrast, capture higher-level intent. They allow site operators to signal permissible or discouraged behaviors in ways that map to human expectations but cannot be reduced to discrete interface elements. Together, these layers provide both fine-grained control and high-level guidance, reducing ambiguity and aligning agent behavior with the preferences of website owners.

\subsection{Resource Rules: Selectors, Verbs, and Modifiers}
\label{sec:resourceRules}

Resource permissions are implemented through \texttt{resource\_rules}, each consisting of a verb, a selector, an allow/deny flag, and optional modifiers. This design reflects several deliberate choices.

\paragraph{Selectors.}
Using CSS selectors leverages well-established, universally supported web technologies. Selectors allow site owners to target arbitrary subsets of a page (e.g., a login form, a navigation menu, or a purchase button) without requiring new annotation standards or custom markup. This choice prioritizes compatibility with existing developer workflows and minimizes deployment costs. Alternatives such as semantic annotations or accessibility trees, by contrast, lack the ubiquity and precision of selector-based targeting.

\paragraph{Verbs.} Supporting multiple verbs acknowledges that interactions with the same element can have different consequences. Reading a content section differs fundamentally from clicking a purchase button or submitting a form. By encoding the action type explicitly, the manifest distinguishes between viewing a resource and interacting with it. Examples of verbs include \texttt{read\_content}, \texttt{click\_element}, \texttt{submit\_form}, \texttt{play\_media}, and so on. This approach is critical for LLM-based agents that perform diverse actions beyond simple crawling.

\paragraph{Modifiers.}
Modifiers enable more nuance in resource rules: rate limits, concurrency caps, time-of-day restrictions, or requiring human-in-the-loop confirmation. These constraints address the fact that harm often emerges from scale rather than intent. A single automated form submission may be acceptable; hundreds per second may overwhelm a server. By codifying these constraints in a structured format, the manifest allows agents to throttle or escalate appropriately, preventing inadvertent denial-of-service while still enabling beneficial automation.

\subsection{Action Guidelines}
\label{sec:actionGuidelines}

Beyond resource rules, the manifest includes \texttt{action\_guidelines}:  semi-structured directives expressed in RFC-2119 terms (MUST, MUST NOT, SHOULD, SHOULD NOT) with natural language descriptions. The rationale is twofold.
First, many high-level behaviors cannot be reduced to element-level rules. A website may wish to prohibit unsolicited direct messages, discourage automated bulk bookings, or require bots to identify themselves in usernames. Encoding such norms at the resource level is infeasible; they are better expressed as action-level guidelines.
Second, natural language descriptions suit the capabilities of LLM-based agents. These models can interpret and incorporate textual rules directly into their reasoning, making natural language an effective channel for defining behavior.

While less easily enforceable than structured rules, action guidelines function as normative signals that can be integrated into the agent's prompt or control logic. This hybrid approach recognizes that both precision (through resource rules) and flexibility (through action guidelines) are necessary for effective interaction policies.

\subsection{APIs as the Preferred Path}
\label{sec:api}

Finally, the manifest allows websites to specify available APIs through the \texttt{api} field, with references to specifications and documentation for OpenAPI, MCP, or A2A endpoints. This reflects a key principle: direct API usage is nearly always preferable to DOM-level interaction. APIs are more efficient, reliable, and auditable, providing stronger guarantees for both parties: websites retain control and observability, while agents gain stable, machine-readable endpoints that reduce breakage and misinterpretation.

However, APIs are not universally available. Many websites lack formal endpoints or expose only partial functionality. The manifest therefore treats APIs as the ``happy path'' but not the sole path. Agents are encouraged to use APIs when possible, but may fall back to DOM-level interactions when necessary. This dual pathway ensures broad applicability without forcing all sites to adopt heavyweight integration efforts.



\section{Analysis}

Having outlined the structure and design principles of \texttt{agent-permissions.json}, we now examine its implications for the broader web ecosystem. This section analyzes the manifest's approach to enforcement, explores the incentive structures that govern its adoption, and evaluates its strengths and limitations. Our analysis demonstrates that the manifest's value lies not in eliminating malicious behavior, but in establishing a compliance framework that makes coordination predictable and straightforward.

\subsection{Enforcement}
\label{sec:enforcement}

A crucial aspect of \texttt{agent-permissions.json} is that, by design, it does not prescribe enforcement mechanisms. This reflects three considerations: (a) mandating strong enforcement would impose infrastructure and operational burdens that many sites, especially smaller ones, cannot meet; (b) confining the standard to permissions keeps the problem tractable and the schema stable; and (c) enforcement technologies and practices (e.g., rate shaping, anomaly detection, attestation, proof-of-work, and tolling) will continue to evolve independently, and decoupling allows that evolution without revising the semantics of permissible behavior.

In practice, the manifest provides an explicit contract. Whatever enforcement a site chooses to layer on top can reference that contract. In other words, the question for the enforcement system shifts from ``is the agent operating with a malicious intent?'' to ``is the agent following the rules in \texttt{agent-permissions.json}?''
This separation enables websites to adopt enforcement strategies appropriate to their resources and threat models while maintaining a stable, universally interpretable permission layer.

\subsection{Incentives}
\label{sec:incentives}

\texttt{agent-permissions.json} is likely to be most useful for websites whose value derives from delivering information or services rather than from human attention. E-commerce platforms, government portals, documentation and research repositories, regulatory filings, scheduling and booking interfaces, and academic repositories all benefit when agents can act within published limits. In these environments, the manifest explicitly invites agent interactions and makes compliance with website policies straightforward.

For instance, many e-commerce platforms actively seek greater indexing and interaction from AI agents associated with platforms such as ChatGPT, as this enhances product visibility in model-generated suggestions and recommendations. Rather than imposing limits, these operators optimize their sites not only according to traditional SEO practices but also with AI agents in mind, structuring content for efficient summarization and exposing dynamic elements for querying.  This emerging ``AI-SEO'' paradigm aligns well with \texttt{agent-permissions.json}, as it allows website owners to signal permissive rules while mitigating abusive patterns.

By contrast, attention- and advertising-funded sites face different incentives: even compliant automation can affect revenue. \texttt{agent-permissions.json} is likely to be less appropriate in such contexts, compared to alternative schemes such as pay-per-interaction, metered access for automated clients, and/or content-use signals such as AIPref for training and derivative use.

From the point of view of agent framework developers, supporting \texttt{agent-permissions.json} is both a gesture of goodwill towards website owners and a pragmatic choice: it offers a low-cost way to reduce the risk of being blocked, rate-limited, or subject to CAPTCHAs, and it provides a clear story about ``compliant mode'' behavior for users and customers. Once permission manifests are widely used, frameworks that ignore them will be harder to deploy on high-value sites, while those that respect them can advertise better reliability and fewer surprises when sites tighten their defenses. The manifest also gives framework authors a single, stable integration point: instead of per-site heuristics or bespoke allowlists/blacklists, they can implement one permission layer that generalizes across the web, much as crawler libraries standardized around \texttt{robots.txt}.

At the same time, the incentives are not entirely one-sided. Since action guidelines are expressed in natural language, a malicious or adversarial site could attempt to use them as a channel for prompt injection or manipulation of agent behavior. Frameworks should therefore treat manifest content as untrusted input and apply the same defenses they already deploy for user- and page-supplied text, such as separating manifest rules from task instructions, applying safety filters, and constraining which parts of the manifest can influence critical actions. In this sense, \texttt{agent-permissions.json} does not introduce a fundamentally new attack surface so much as it formalizes one more source of instructions that an agent must reason about and cross-check against its own policies, operator constraints, and user intent.

\subsection{Strengths and Limitations}

\texttt{agent-permissions.json} inherits many of its strengths from the philosophy that made \texttt{robots.txt} a durable standard. Its first strength is simplicity: it requires no specialized infrastructure, authentication backends, or cryptographic identity schemes. A manifest file can be published at a well-known location, updated as needed, and read by any compliant agent. This simplicity is not merely a convenience; it is the basis for broad adoption, since the vast majority of websites, particularly smaller operators, lack the resources to deploy heavyweight solutions.

Second, the manifest is low-friction for agents. Implementing compliance requires only basic parsing and respect for declarative rules, which can be incorporated into agent frameworks with negligible overhead. In practice, the key form of ``support'' is from the maintainers of widely used agent stacks: headless-browser drivers, orchestration frameworks, and hosted agent platforms. Once these libraries ship manifest parsing and checks as part of their default execution path, the path of least resistance for deployers is to leave those checks enabled. Ignoring \texttt{agent-permissions.json} then becomes an explicit opt-out decision that requires additional configuration and, for commercial providers, carries reputational and potentially legal downside.

This does not eliminate the possibility that some users will instruct their own agents to disregard permissions, especially in self-hosted or research settings. However, large-scale and public-facing deployments are unlikely to normalize such behavior: they benefit from being seen as compliant clients (for example, by being less likely to be blocked or rate-limited) and typically operate under internal policies that disallow deliberate circumvention of published rules. 
In this sense, widespread support makes ``respecting permissions'' the default in the same way that honoring \texttt{robots.txt} has become the default for mainstream crawlers: standard libraries consult the manifest automatically and enforce disallow rules, rate limits, and human-in-the-loop modifiers unless they are deliberately disabled. 


Finally, the manifest facilitates a compliance mechanism that can evolve gradually. Just as \texttt{robots.txt} grew from simple allow/disallow directives to support features like crawl-delay, \texttt{agent-permissions.json} is explicitly designed to admit richer policy constructs over time without breaking existing clients. The JSON representation, together with an explicit \texttt{schema\_version} in the \texttt{metadata} block, allows new fields (for example, additional verbs, modifiers, or action guideline types) to be introduced as optional extensions that compliant agents may ignore if unknown. This means that early adopters can rely on a small, stable core of semantics, while more sophisticated sites and frameworks can progressively experiment with finer-grained controls (such as per-user classes, time-of-day rules, or richer human-in-the-loop requirements). 



Despite these strengths, the manifest also inherits limitations from \texttt{robots.txt}. Most significantly, it is non-enforceable by design: it functions as a signaling and coordination tool rather than as an enforcement mechanism. Nothing prevents a malicious agent from disregarding the file entirely. However, as discussed in \Cref{sec:enforcement}, this limitation is also a strength: by avoiding commitments to enforcement infrastructure, the manifest remains lightweight, general, and future-proof. Enforcement can be layered on top through complementary mechanisms. In this regard, the manifest serves as a foundational layer: a minimal, universally applicable mechanism upon which more robust enforcement and accountability structures can be built.


In summary, \texttt{agent-permissions.json} represents a shift from blanket blocking to compliance-based coordination.
It does not eliminate malicious behavior, but it does reduce collateral damage by giving compliant agents a clear path to cooperation and enabling website owners to coordinate with cooperative agents. Its strength lies in its modesty: by doing one thing simply and effectively, i.e., making permissions explicit, it creates the conditions for a healthier ecosystem of human, agent, and website interactions.

\section{Implementation}

The \texttt{agent-permissions.json} manifest is deliberately lightweight, so its value depends on practical support from both agent frameworks (e.g., LangChain, CrewAI, Camel) and website operators. In this section, we discuss how \texttt{agent-permissions.json} can be integrated into agent frameworks and web stacks. To demonstrate the feasibility of these integrations, we describe two reference implementations we developed: a Python library for agents and a manifest generator tool for site owners. 

Taken together, these reference implementations demonstrate that the manifest is practical for both agents and websites. For agents, resource rules can be deterministically enforced at the browser layer, action guidelines can be incorporated with available information, and API references can be integrated into planning logic. For websites, manifest creation can be as simple as publishing a static file or an automated build-system integration for more complex deployments. These prototypes confirm that the standard achieves its goal of enabling structured coordination with minimal implementation overhead, paving the way for broader adoption.

\subsection{Agent Frameworks}

For automated agents, following manifest rules aligns naturally with the three layers of the specification.

\paragraph{Resource Rules.}
Resource-level directives are the most straightforward to follow. Since most agents rely on headless browser automation (e.g., Selenium, Playwright, and Puppeteer) to interact with websites, we can apply resource rules at the browser interface layer. The mechanism is simple: when the agent attempts an interaction, the driver checks the manifest before issuing the corresponding DOM command. If the action is disallowed, the attempt is blocked. This ensures that prohibited interactions never reach the page, creating a deterministic enforcement boundary.

\paragraph{Action Guidelines.}
Action guidelines are inherently less deterministic. They are expressed in semi-structured natural language and are intended to be incorporated into the agent's reasoning process rather than enforced mechanically. In practice, they can be provided to the agent's control policy as normative signals that shape behavior. While this cannot guarantee compliance, the lack of enforceability is not unique to this standard; it reflects the general challenge of constraining LLM behavior with natural language. Our reference implementation provides action rules as structured prompt material, enabling agents to integrate them into their deliberation loop.

It is important to note that, as with all LLM signaling tools and LLM-readable interfaces, natural language guidelines can be a vector for prompt injection attacks  \citep[similar to UIs designed to manipulate agents;][]{aichberger2025attacking,wang2025webinject}. As we discussed in \Cref{sec:incentives}, action guidelines should therefore be treated with the same security policies as any other element with which the agent interacts.

\paragraph{API Specifications} The \texttt{api} field indicates preferred endpoints that agents should use instead of DOM-level interaction. These are advisory rather than binding: agents are encouraged to prefer API calls when available, but fallback to resource-level interactions remains permissible if no corresponding API exists. The rationale is pragmatic: APIs are more reliable and efficient but not universally available. At the implementation level, the agent framework exposes the manifest's API references as part of its planning logic, allowing agents to prioritize them when constructing execution strategies.

\paragraph{Reference Library.} To facilitate adoption, we developed a Python library\footnote{\url{https://github.com/las-wg/agent-permissions-python}} that parses \texttt{agent-permissions.json} files, validates them against the schema, and exposes their contents in machine-readable form for the agent framework. The library handles caching and schema validation, as well as providing a thin abstraction layer for integration with headless browser drivers. For example, a rule disallowing form submissions can be enforced by wrapping the driver's \texttt{submit()} call with a manifest check. Similarly, modifiers such as rate limits or human-in-the-loop requirements can be surfaced to the agent framework as constraints, enabling compliance without significant additional logic. This library shows that manifest integration is feasible with minimal engineering effort while providing a concrete starting point for broader ecosystem support.

\subsection{Website Owners}

On the publishing side, the implementation burden is intentionally minimal: an \texttt{agent-permissions.json} file must simply be placed in \texttt{./well-known} and updated as needed. To further reduce friction and encourage adoption, we explored how to streamline the creation of manifests.

Specifically, we developed a tool\footnote{\url{https://github.com/las-wg/agent-permissions-generator}} that generates manifest files automatically by crawling a website and applying developer-specified rules. Given a set of natural language policies (e.g., ``don't let agents register an account'', ``don't let agents click links with `buy' in the text'', ``allow agents to send messages''), the tool traverses the site's DOM, identifies relevant selectors, and emits a valid \texttt{agent-permissions.json} file with the appropriate resource and action rules. This allows developers to define rules declaratively without manually inspecting each page or writing selectors by hand. The tool could be extended and integrated into build systems, continuous integration pipelines, or content management systems, ensuring that the manifest is automatically generated and kept updated as the site evolves.

This dual approach (minimal manual requirements for small sites, coupled with automated generation for larger or more dynamic ones) means that the barrier to adoption is low. Website operators can choose between hand-crafted manifests for precision or automated generation for scalability, depending on their needs and resources. Crucially, the manifest integrates smoothly with existing development practices, requiring no changes to authentication systems, back-end logic, or page architecture.

\section{Conclusion}

The automated web has shifted from a mostly read-only setting to one where LLM-based agents routinely navigate interfaces, make complex decisions, and execute workflows. This proliferation of agentic interaction has outpaced existing governance mechanisms and led to increasingly defensive responses: blanket blocking, aggressive CAPTCHAs, and indiscriminate rate limiting that treat all automated traffic as equally risky.

We have introduced \texttt{agent-permissions.json}, a minimal, declarative layer that lets websites state what forms of automated interaction are acceptable and on what terms. By distinguishing between resource-level rules (deterministically enforceable at the DOM or browser layer) and action-level guidelines (integrated into agent reasoning), the standard provides both precision and flexibility. Critically, it requires no authentication infrastructure or complex middleware: just a static file that agents can discover and follow. The design draws inspiration from \texttt{robots.txt} while addressing the richer interaction patterns of modern agents.

In practice, the manifest does not attempt to eliminate malicious behavior. Instead, it establishes a stable compliance framework that makes cooperation predictable: compliant agents have a clear path to behaving correctly, and website operators have a standard format to specify policies. This shifts the question for enforcement systems from ``is this client automated?'' to ``is this client respecting the published rules?''

Of course, the standard's impact will depend on adoption. Agent frameworks and hosted platforms must integrate manifest parsing and enforcement into their default execution paths, and website operators must publish and maintain manifests that reflect their preferences. Our reference implementations (a Python library for enforcing resource rules and a generator that emits manifests from developer-specified policies) demonstrate that this integration is feasible with modest engineering effort.

\texttt{agent-permissions.json} is not a complete solution to all challenges posed by AI agents on the web, nor is it a substitute for content-use standards, economic mechanisms, or security tooling. Instead, it provides a foundation: a common, lightweight language for cooperative behavior. By making permissions explicit, it creates conditions for a healthier ecosystem in which websites can invite beneficial automation on their own terms, agents can reliably act in a beneficial manner, and enforcement can be focused where it belongs: on those who choose to ignore the rules.

\section*{Acknowledgements}

We would like to thank Leslie Tao and Emanuele La Malfa for their support. Samuele Marro is supported by the EPSRC Centre for Doctoral Training in Autonomous Intelligent
Machines and Systems n. EP/Y035070/1, in addition to Microsoft Ltd.

\bibliographystyle{tmlr}
\bibliography{tmlr}

@article{nakano2021webgpt,
  title={Webgpt: Browser-assisted question-answering with human feedback},
  author={Nakano, Reiichiro and Hilton, Jacob and Balaji, Suchir and Wu, Jeff and Ouyang, Long and Kim, Christina and Hesse, Christopher and Jain, Shantanu and Kosaraju, Vineet and Saunders, William and others},
  journal={arXiv preprint arXiv:2112.09332},
  year={2021}
}

@article{drouin2024workarena,
  title={Workarena: How capable are web agents at solving common knowledge work tasks?},
  author={Drouin, Alexandre and Gasse, Maxime and Caccia, Massimo and Laradji, Issam H and Del Verme, Manuel and Marty, Tom and Boisvert, L{\'e}o and Thakkar, Megh and Cappart, Quentin and Vazquez, David and others},
  journal={arXiv preprint arXiv:2403.07718},
  year={2024}
}

@article{baek2024researchagent,
  title={Researchagent: Iterative research idea generation over scientific literature with large language models},
  author={Baek, Jinheon and Jauhar, Sujay Kumar and Cucerzan, Silviu and Hwang, Sung Ju},
  journal={arXiv preprint arXiv:2404.07738},
  year={2024}
}

@inproceedings{kodandaram2024enabling,
  title={Enabling uniform computer interaction experience for blind users through large language models},
  author={Kodandaram, Satwik Ram and Uckun, Utku and Bi, Xiaojun and Ramakrishnan, IV and Ashok, Vikas},
  booktitle={Proceedings of the 26th International ACM SIGACCESS Conference on Computers and Accessibility},
  pages={1--14},
  year={2024}
}

@misc{bocharov2024declare,
  author       = {Alex Bocharov and Santiago Vargas and Adam Martinetti and Reid Tatoris and Carlos Azevedo},
  title        = {Declare your {AIndependence}: block AI bots, scrapers and crawlers with a single click},
  year         = {2024},
  month        = {jul},
  day          = {3},
  url          = {https://blog.cloudflare.com/declaring-your-aindependence-block-ai-bots-scrapers-and-crawlers-with-a-single-click}
}

@misc{corral2025perplexity,
  author = {Gabriel Corral and Vaibhav Singhal and Brian Mitchell and Reid Tatoris},
  title  = {Perplexity is using stealth, undeclared crawlers to evade website no-crawl directives},
  year   = {2025},
  month  = {aug},
  day    = {4},
  url    = {https://blog.cloudflare.com/perplexity-is-using-stealth-undeclared-crawlers-to-evade-website-no-crawl-directives/},
  note   = {Cloudflare Blog}
}

@techreport{w3c2021inaccessibility,
  author      = {Scott Hollier and Janina Sajka and Jason White and Michael Cooper},
  title       = {Inaccessibility of CAPTCHA: Alternatives to Visual Turing Tests on the Web},
  type        = {W3C Group Draft Note},
  institution = {World Wide Web Consortium (W3C)},
  year        = {2021},
  month       = {dec},
  day         = {16},
  number      = {DNOTE-turingtest-20211216},
  url         = {https://www.w3.org/TR/turingtest/}
}

@techreport{fletcher2024many,
  title={How many news websites block AI crawlers?},
  author={Fletcher, Richard},
  year={2024},
  institution={Reuters Institute for the Study of Journalism}
}

@article{llamas2025balancing,
  title={Balancing Security and Privacy: Web Bot Detection, Privacy Challenges, and Regulatory Compliance under the GDPR and AI Act},
  author={Llamas, Javier Mart{\'\i}nez and Vranckaert, Koen and Preuveneers, Davy and Joosen, Wouter},
  journal={Open Research Europe},
  volume={5},
  pages={76},
  year={2025}
}

@article{iliou2021detection,
  title={Detection of advanced web bots by combining web logs with mouse behavioural biometrics},
  author={Iliou, Christos and Kostoulas, Theodoros and Tsikrika, Theodora and Katos, Vasilis and Vrochidis, Stefanos and Kompatsiaris, Ioannis},
  journal={Digital Threats: Research and Practice},
  volume={2},
  number={3},
  pages={1--26},
  year={2021},
  publisher={ACM New York, NY, USA}
}

@article{reuters2025reddit,
  author       = {{Reuters}},
  title        = {Reddit sues AI startup Anthropic, allegedly using data without permission},
  year         = {2025},
  month        = {jun},
  day          = {4},
  journal      = {Reuters},
  url           = {https://www.reuters.com/business/reddit-sues-ai-startup-anthropic-allegedly-using-data-without-permission-2025-06-04/}
}

@article{pierce2024text,
  author      = {David Pierce},
  title       = {The text file that runs the internet},
  year        = {2024},
  month       = {feb},
  day         = {14},
  journal     = {The Verge},
  url         = {https://www.theverge.com/24067997/robots-txt-ai-text-file-web-crawlers-spiders},
  note        = {Accessed: 2025-10-13}
}

@article{kim2025scrapers,
  title={Scrapers selectively respect robots. txt directives: evidence from a large-scale empirical study},
  author={Kim, Taein and Bock, Karstan and Luo, Claire and Liswood, Amanda and Wenger, Emily},
  journal={arXiv preprint arXiv:2505.21733},
  year={2025}
}

@article{reitinger2025measured,
  title={Measured Failures of robots. txt: Legal and Empirical Insights for Regulating Robots},
  author={Reitinger, Nathan},
  journal={Northwestern Public Law Research Paper},
  number={25-53},
  year={2025}
}

@article{li2025ai,
  title={ai. txt: A Domain-Specific Language for Guiding AI Interactions with the Internet},
  author={Li, Yuekang and Song, Wei and Zhu, Bangshuo and Gong, Dong and Liu, Yi and Deng, Gelei and Chen, Chunyang and Ma, Lei and Sun, Jun and Walsh, Toby and others},
  journal={arXiv preprint arXiv:2505.07834},
  year={2025}
}

@techreport{rfc9309,
  author      = {Martijn Koster and Gary Illyes and Henner Zeller and Lizzi Sassman},
  title       = {Robots Exclusion Protocol},
  type        = {RFC},
  number      = {9309},
  institution = {Internet Engineering Task Force (IETF)},
  year        = {2022},
  month       = {sep},
  doi         = {10.17487/RFC9309},
  url         = {https://www.rfc-editor.org/rfc/rfc9309.html}
}

@inproceedings{sun2012devil,
  title={The devil is in the (implementation) details: an empirical analysis of OAuth SSO systems},
  author={Sun, San-Tsai and Beznosov, Konstantin},
  booktitle={Proceedings of the 2012 ACM conference on Computer and communications security},
  pages={378--390},
  year={2012}
}

@inproceedings{yang2016model,
  title={Model-based security testing: An empirical study on oauth 2.0 implementations},
  author={Yang, Ronghai and Li, Guanchen and Lau, Wing Cheong and Zhang, Kehuan and Hu, Pili},
  booktitle={Proceedings of the 11th ACM on Asia Conference on Computer and Communications Security},
  pages={651--662},
  year={2016}
}

@article{olston2010web,
  title={Web crawling},
  author={Olston, Christopher and Najork, Marc and others},
  journal={Foundations and Trends{\textregistered} in Information Retrieval},
  volume={4},
  number={3},
  pages={175--246},
  year={2010},
  publisher={Now Publishers, Inc.}
}

@article{trezza2023scrape,
  title={To scrape or not to scrape, this is dilemma. The post-API scenario and implications on digital research},
  author={Trezza, Domenico},
  journal={Frontiers in sociology},
  volume={8},
  pages={1145038},
  year={2023},
  publisher={Frontiers Media SA}
}

@article{brown2024web,
  title={Web scraping for research: Legal, ethical, institutional, and scientific considerations},
  author={Brown, Megan A and Gruen, Andrew and Maldoff, Gabe and Messing, Solomon and Sanderson, Zeve and Zimmer, Michael},
  journal={arXiv preprint arXiv:2410.23432},
  year={2024}
}

@misc{rfc6749,
  series     = {Request for Comments},
  number     = {6749},
  howpublished = {RFC 6749},
  publisher  = {RFC Editor},
  author     = {Dick Hardt},
  title      = {The OAuth 2.0 Authorization Framework},
  year       = {2012},
  month      = {oct},
  doi        = {10.17487/RFC6749},
  url        = {https://www.rfc-editor.org/info/rfc6749}
}

@techreport{nistsp80057,
  author      = {Barker, Elaine},
  title       = {Recommendation for Key Management: Part 1 -- General},
  institution = {National Institute of Standards and Technology},
  type        = {NIST Special Publication},
  number      = {800-57 Part 1, Revision 5},
  year        = {2020},
  month       = {may},
  doi         = {10.6028/NIST.SP.800-57pt1r5},
  url         = {https://csrc.nist.gov/pubs/sp/800/57/pt1/r5/final}
}

@misc{googlecloudApi,
  author  = {{Google Cloud}},
  title   = {Best practices for managing API keys},
  url     = {https://cloud.google.com/docs/authentication/api-keys-best-practices},
  note    = {Google Cloud Documentation},
  year    = {2025},
  urldate = {2025-10-14}
}

@misc{owaspfoundation2025key,
  author  = {{OWASP Foundation}},
  title   = {Key Management Cheat Sheet},
  year    = {2025},
  url     = {https://cheatsheetseries.owasp.org/cheatsheets/Key_Management_Cheat_Sheet.html},
  note    = {OWASP Cheat Sheet Series},
  urldate = {2025-10-14}
}

@inproceedings{prandl2015study,
  title={A study of web application firewall solutions},
  author={Prandl, Stefan and Lazarescu, Mihai and Pham, Duc-Son},
  booktitle={International conference on information systems security},
  pages={501--510},
  year={2015},
  organization={Springer}
}

@article{zhou2023webarena,
  title={Webarena: A realistic web environment for building autonomous agents},
  author={Zhou, Shuyan and Xu, Frank F and Zhu, Hao and Zhou, Xuhui and Lo, Robert and Sridhar, Abishek and Cheng, Xianyi and Ou, Tianyue and Bisk, Yonatan and Fried, Daniel and others},
  journal={arXiv preprint arXiv:2307.13854},
  year={2023}
}

@article{bort2025how,
  author  = {Julie Bort},
  title   = {How OpenAI’s bot crushed this seven-person company’s website ‘like a DDoS attack’},
  journal = {TechCrunch},
  year    = {2025},
  month   = {jan},
  day     = {10},
  url     = {https://techcrunch.com/2025/01/10/how-openais-bot-crushed-this-seven-person-companys-web-site-like-a-ddos-attack/}
}

@article{paul2024multiple,
  author  = {Katie Paul},
  title   = {Multiple AI companies bypassing web standard to scrape publisher sites, licensing firm says},
  journal = {Reuters},
  year    = {2024},
  month   = {jun},
  day     = {21},
  url     = {https://www.reuters.com/technology/artificial-intelligence/multiple-ai-companies-bypassing-web-standard-scrape-publisher-sites-licensing-2024-06-21}
}

@article{zhang2024llasa,
  title={Llasa: Large language and e-commerce shopping assistant},
  author={Zhang, Shuo and Peng, Boci and Zhao, Xinping and Hu, Boren and Zhu, Yun and Zeng, Yanjia and Hu, Xuming},
  journal={arXiv preprint arXiv:2408.02006},
  year={2024}
}

@article{lazar2025moral,
  title={The moral case for using language model agents for recommendation},
  author={Lazar, Seth and Thorburn, Luke and Jin, Tian and Belli, Luca},
  journal={Inquiry},
  pages={1--25},
  year={2025},
  publisher={Taylor \& Francis}
}

@article{south2025authenticated,
  title={Authenticated delegation and authorized ai agents},
  author={South, Tobin and Marro, Samuele and Hardjono, Thomas and Mahari, Robert and Whitney, Cedric Deslandes and Greenwood, Dazza and Chan, Alan and Pentland, Alex},
  journal={arXiv preprint arXiv:2501.09674},
  year={2025}
}

@article{boisvert2024workarenapp,
  title={Workarena++: Towards compositional planning and reasoning-based common knowledge work tasks},
  author={Boisvert, L{\'e}o and Thakkar, Megh and Gasse, Maxime and Caccia, Massimo and de Chezelles, Thibault and Cappart, Quentin and Chapados, Nicolas and Lacoste, Alexandre and Drouin, Alexandre},
  journal={Advances in Neural Information Processing Systems},
  volume={37},
  pages={5996--6051},
  year={2024}
}

@article{erdogan2025plan,
  title={Plan-and-act: Improving planning of agents for long-horizon tasks},
  author={Erdogan, Lutfi Eren and Lee, Nicholas and Kim, Sehoon and Moon, Suhong and Furuta, Hiroki and Anumanchipalli, Gopala and Keutzer, Kurt and Gholami, Amir},
  journal={arXiv preprint arXiv:2503.09572},
  year={2025}
}

@article{wang2024oscar,
  title={Oscar: Operating system control via state-aware reasoning and re-planning},
  author={Wang, Xiaoqiang and Liu, Bang},
  journal={arXiv preprint arXiv:2410.18963},
  year={2024}
}

@misc{howard2024llmstxt,
  author       = {Jeremy Howard},
  title        = {The /llms.txt file},
  year         = {2024},
  organization = {llms-txt},
  howpublished = {\url{https://llmstxt.org/}},
  note         = {Published September 3, 2024. Accessed November 8, 2025.}
}

@misc{cloudflare2025paypercrawl,
  author    = {Will Allen and Simon Newton},
  title     = {Introducing Pay Per Crawl: Enabling Content Owners to Charge AI Crawlers for Access},
  year      = {2025},
  month     = {7},
  date      = {2025-07-01},
  url       = {https://blog.cloudflare.com/introducing-pay-per-crawl/},
  publisher = {Cloudflare},
  urldate   = {2025-11-08}
}

@article{aichberger2025attacking,
  title={Attacking multimodal os agents with malicious image patches},
  author={Aichberger, Lukas and Paren, Alasdair and Gal, Yarin and Torr, Philip and Bibi, Adel},
  journal={arXiv preprint arXiv:2503.10809},
  year={2025}
}

@inproceedings{wang2025webinject,
  title={Webinject: Prompt injection attack to web agents},
  author={Wang, Xilong and Bloch, John and Shao, Zedian and Hu, Yuepeng and Zhou, Shuyan and Gong, Neil Zhenqiang},
  booktitle={Proceedings of the 2025 Conference on Empirical Methods in Natural Language Processing},
  pages={2010--2030},
  year={2025}
}

@misc{mcp,
  title        = {Model Context Protocol (MCP) Specification},
  author       = {{Anthropic}},
  year         = {2025},
  month        = {6},
  date         = {2025-06-18},
  url          = {https://modelcontextprotocol.io/specification/2025-06-18},
  urldate      = {2025-11-09},
  note         = {Official specification}
}

@misc{a2a,
  title        = {A2A Protocol (Agent-to-Agent)},
  author       = {{Google}},
  year         = {2025},
  url          = {https://a2a-protocol.org/},
  urldate      = {2025-11-09},
  note         = {Agent-to-agent interoperability standard}
}

@misc{x402,
  title        = {MCP Server with x402},
  author       = {{Coinbase}},
  year         = {2025},
  url          = {https://docs.cdp.coinbase.com/x402/mcp-server},
  urldate      = {2025-11-09},
  note         = {Using the x402 payment protocol with MCP}
}

@misc{ap2,
  title        = {Announcing Agent Payments Protocol (AP2)},
  author       = {{Google}},
  year         = {2025},
  month        = {9},
  date         = {2025-09-16},
  url          = {https://cloud.google.com/blog/products/ai-machine-learning/announcing-agents-to-payments-ap2-protocol},
  urldate      = {2025-11-09},
  note         = {Open protocol for agent-led payments}
}

@misc{robots2025,
  author       = {Martin Splitt and John Mueller},
  title        = {Robots Refresher: robots.txt --- a flexible way to control how machines explore your website},
  year         = {2025},
  month        = mar,
  day          = {7},
  url          = {https://developers.google.com/search/blog/2025/03/robotstxt-flexible-way-to-control},
  urldate      = {2025-11-29},
  organization = {Google Search Central Blog, Google for Developers}
}

@misc{cloudflareWebCrawler,
  author  = {{Cloudflare}},
  title   = {What is a web crawler? {How} web spiders work},
  year    = {2025},
  url     = {https://www.cloudflare.com/learning/bots/what-is-a-web-crawler},
  note    = {Accessed: 2025-11-29}
}


\end{document}